\documentclass[aps,prd,showpacs,groupedaddress,nofootinbib]{revtex4}

\usepackage{color}
\usepackage{graphicx}
\usepackage{amsmath}
\begin{document}

\title{Biomechanical conditions of walking}

\author{Y. F. Fan$^{1}$\footnote{Corresponding authors\\ Email: tfyf@fjnu.edu.cn}, L. P. Luo$^{2}$, Z. Y. Li$^{1}$, S. Y. Han$^{1}$, C. S. Lv$^{3}$ and B. Zhang$^{3}$}
\affiliation{
$^{1}$School of Physical Education and Sport Science, Fujian Normal University, Fuzhou, 350117, P.R. China\\
$^{2}$Medical Imaging Center, the First Affiliated Hospital of Jinan University, Guangzhou 510632, P.R. China\\
$^{3}$Center for Scientific Research, Guangzhou Institute of Physical Education, Guangzhou 510500, P.R. China}

\date{\today}

\begin{abstract}
The development of rehabilitation training program for lower limb injury does not usually include gait pattern design. This paper introduced a gait pattern design by using equations (conditions of walking). Following the requirements of reducing force to the injured side to avoid further injury, we developed a lower limb gait pattern to shorten the stride length so as to reduce walking speed, to delay the stance phase of the uninjured side and to reduce step length of the uninjured side. This gait pattern was then verified by the practice of a rehabilitation training of an Achilles tendon rupture patient, whose two-year rehabilitation training (with 24 tests) has proven that this pattern worked as intended. This indicates that rehabilitation training program for lower limb injury can rest on biomechanical conditions of walking based on experimental evidence.

\end{abstract}


\maketitle

\section{Introduction}
Gait patterns are described by gait parameters such as speed, step length and cadence \cite{1}. Though gait parameters of each individual vary \cite{2,3}, healthy people¡¯s walking features of symmetry and economy are consistent \cite{4,5}. The inverted pendulum model typically simulates these features, and gait pattern equations were established \cite{6,7,8}. Gait patterns vary when their body dysfunctions \cite{9}. When walking is prescribed as part of rehabilitation training program \cite{10}, it is necessary to design gait patterns to avoid further injury \cite{11}. But the designs of gait pattern are mostly concerned with walking speed control at early rehabilitation period \cite{12}.

Lower limb joints of ankle, knee and hip get injured easily and frequently. Training exercise is effective for such injury rehabilitation. For example, at early stage of Achilles tendon rupture (ATR) rehabilitation, controlled tensional loading to the end of tendon rupture can improve the biomechanical properties of scar tissue, reduce adhesions, stimulate fibroblast collagen fibers, and improve mechanical properties after ATR recovery. Early training can yield positive results \cite{13,14}. But not many gait patterns have been developed and included in early rehabilitation period, leading to a lack of agreed effective methods \cite{13,15}.

When injury happens at one side, that side is called the injured side and the other uninjured side. Basic early rehabilitation training requirements have been agreed: to reduce force to the injured side to avoid further injury. During walking stance phase, forces between lower limb and support surface interact. To the case of one side injury, we hypothesized that when walking, the peak values of vertical ground reaction force (VGRF) of the injured side could be reduced to be close to the weight by changing gait patterns. Gait patterns were simulated by equations. The results demonstrated that these patterns were effective. One ATR patient adopted our gait pattern. In the following two years, we took 24 tests. The results verified our hypothesis.

\section{Gait pattern design}

\subsection{Vertical ground reaction force}
$F_{I}$ was set as VGRF of the injured side, $F_{C}$ as that of the uninjured side, $m$ as weight, $T$ as stride time. For walking, these quantities should meet conditions from the following equation:
\begin{equation}
\label{eqn-1}
\int^{T}_{0}\left( F_{I}+F_{C}\right)dt=\int^{T}_{0}mgdt
\end{equation}

\begin{figure}[!ht]
\begin{center}
\begin{tabular}{cccc}
 \includegraphics[width=14.8cm]{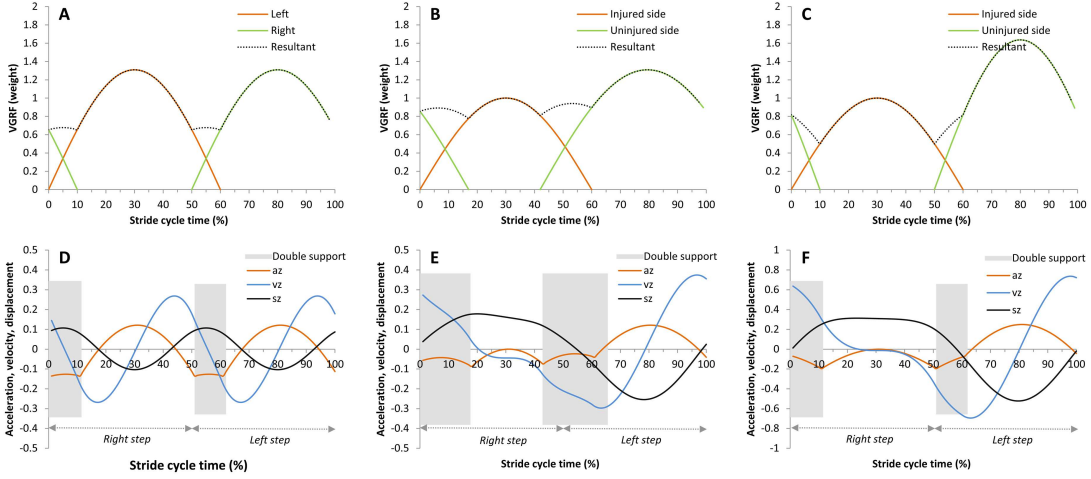}
\end{tabular}
\caption{\label{fig1} Gait pattern design. A Healthy people's VGRF distribution. B VGRF distribution of the uninjured side when it delayed the stance phase after the other side got injured. C VGRF distribution of the uninjured side when it increased its stretching time after the other side got injured. D Healthy people's acceleration, speed and displacement of COM. E Delay of the stance phase of the uninjured side. F Increase of VGRF of the uninjured side. Both sides' VGRF were simplified as selected function. When the percentage of the stance phase of the uninjured side was 60\%, the peak value of VGRF was 1.309 times more than the weight; when the percentage was 75\%, the peak value of VGRF was still 1.309 times, but that of the injured side fell to be close to the weight; when the peak value of VGRF of the uninjured side was increased to 1.637 times more than the weight, the peak values of VGRF of the injured side fell to be close to the weight.}
\end{center}
\end{figure}

When people walk, the VGRF forms a double-hump pattern, and the hump values exceed the weight. To meet the conditions of reducing force to the injured side and avoiding further injury, Eq.~(\ref{eqn-1}) presented two methods to reduce peak values of VGRF of the injured side: to increase $F_{C}$, or to delay the stance phase of the uninjured side, because $\int^{T}_{0}F_{C}dt=\int^{T_{s}}_{0}F_{C}dt$ ($T=T_{s}+T_{w}$, $\int^{T_{w}}_{0}F_{C}dt=0$, where $T_{s}$ was stance phase, and $T_{w}$ swing phase). Under these conditions, inverse kinematics \cite{16,17} was used to calculate acceleration, speed and displacement of the center of mass (COM). See Fig.~\ref{fig1} for calculation results.

Fig.~\ref{fig1} compared these two methods to increase VGRF or to delay the stance phase of the uninjured side. We found that in a stride cycle, the speed of COM varied ¨C the difference between the maximal and minimal value was 113.13\% ($\left(\left(v^{max}_{1}-v^{min}_{1}\right)-\left(v^{max}_{2}-v^{min}_{2}\right)\right)/\left(v^{max}_{2}-v^{min}_{2}\right)\times 100$, subscript 1 referring to a method to increase $F_{C}$ of the uninjured side, subscript 2 to a method to delay the stance phase of the uninjured side, and $v$ to speed of COM); the difference between the maximal and minimal value of the displacement of COM was 93.12\% ($\left(\left(d^{max}_{1}-d^{min}_{1}\right)-\left(d^{max}_{2}-d^{min}_{2}\right)\right)/\left(d^{max}_{2}-d^{min}_{2}\right)\times 100$, subscript 1 referring to a method to increase $F_{C}$ of the uninjured side, subscript 2 to a method to delay the stance phase of the uninjured side, and $d$ to displacement of COM). As we know, when walking, the greater the changes of speed and displacement in vertical direction, the less stable the walking, and the more risk to the injured side. A more desirable gait pattern for the injured lower limb should be: to delay the stance phase of the uninjured side.

\subsection{Stride time}
$T_{C-I}$ was set as the first double support of the uninjured side, $T_{I-C}$ the second double support of the uninjured side, $T_{I}$ swing phase of the injured side, i.e. single support of the uninjured side, $T_{C}$ swing phase of the uninjured side, i.e. single support of the injured side, $T^{S}_{I}$ stride time of the injured side, and $T^{S}_{C}$ stride time of the uninjured side. For walking training, these quantities should meet conditions from the following equation:

\begin{equation}
\label{eqn-2}\left\{
\begin{array}{c}
T^{S}_{I}=T^{S}_{C}\\
T_{C-I}+T_{I}+T_{I-C}+T_{C}=T^{S}_{C}\\
\end{array}\right.
\end{equation}

Eq.~(\ref{eqn-2}) presented two methods to increase the stance phase of the uninjured side: to only increase $T_{C-I}$ and $T_{I-C}$; or to increase $T_{C-I}$ and $T_{I-C}$, and to reduce $T_{C}$. If we only increased $T_{C-I}$ and $T_{I-C}$, it might increase $T_{I}$, which might bring pitfall to the injured side. Therefore, gait pattern for the injured lower limb should be: to increase $T_{C-I}$ and $T_{I-C}$, and at the same time to reduce $T_{C}$ to increase stance phase of the uninjured side, and to reduce single support of the uninjured side.

\subsection{Stride length}
$L_{C}$ was set as step length of the uninjured side (the length between the heels from the injured side's heel strike to the uninjured side's heel strike on the same side), $L_{I}$ as step length of the injured side (the length between the heels from the uninjured side's heel strike to the injured side's heel strike on the same side), $L^{S}_{C}$ as stride length of the uninjured side, and $L^{S}_{I}$ as stride length of the injured side. For walking training, these quantities should meet conditions from the following equation:

\begin{equation}
\label{eqn-3}\left\{
\begin{array}{c}
L^{S}_{I}=L^{S}_{C}\\
L_{C}+L_{I}=L^{S}_{C}\\
L_{I}+L_{C}=L^{S}_{I}\\
\end{array}\right.
\end{equation}	
	
Eq.~(\ref{eqn-3}) presented two methods to change the distance between the injured and uninjured side: to reduce the step length between the uninjured and injured side (i.e. to increase the step length between the injured and uninjured side); or to increase the step length between the uninjured and injured side (i.e. to shorten the step length between the injured and uninjured side). To make the uninjured side support the injured side earlier, it was more advisable to reduce $L_{C}$.

\subsection{Speed}
$V$ was set as speed, $f$ as cadence. For walking, these quantities and stride length should meet conditions of the following equation:
\begin{equation}
\label{eqn-4}
V=L^{S}_{I}f=L^{S}_{C}f=(L_{I}+L_{C})f=(L_{C}+L_{I})f
\end{equation}

Eq.~(\ref{eqn-4}) presented two methods to control speed: to reduce cadence or to reduce stride length. When walking, VGRF determined not only changes of speed and displacement of COM at the vertical direction, but it also determined stride length because the stride length was determined by friction, while friction was related to the VGRF. To reduce VGRF meant to reduce stride length. Therefore, it was more advisable to reduce stride length so as to slow down speed.

Based upon Eq.~(\ref{eqn-1})$-$(4), gait pattern for the rehabilitation training program of lower limb single side injury should be designed as follows:
1) To delay stance phase from the uninjured side; 2) to reduce step length from the uninjured side; and 3) to slow down speed by reducing stride length.

\section{Experimental verification}
To examine whether the gait pattern designed for the early rehabilitation training of lower limb injury by using Equations 1-4 is scientific, we will test it.

Participant: One 50-year-old participant had an ATR when he was playing badminton without wearing badminton shoes or doing warm-up exercises in a rainy day. When he was trying to step forward on his right side to catch the ball, he felt a sudden heavy hit on his left Achilles tendon. His left foot could not support the weight, so he sat on the floor. In the hospital, the doctor diagnosed it as acute ATR, and performed a surgery the same day. After the surgery, he had to wear a cast. After he had his cast removed, he began to do rehabilitation training to stand and walk on his injured side, using assistive devices such as a walking stick. The patient and his doctor agreed to adopt our designed gait patterns. We began to do gait tests in the laboratory three days after his rehabilitation training began. In the following 98 weeks (07/2012 to 05/2014), he took 24 tests.

\begin{figure}[!ht]
\begin{center}
\begin{tabular}{cccc}
\includegraphics[width=14.8cm]{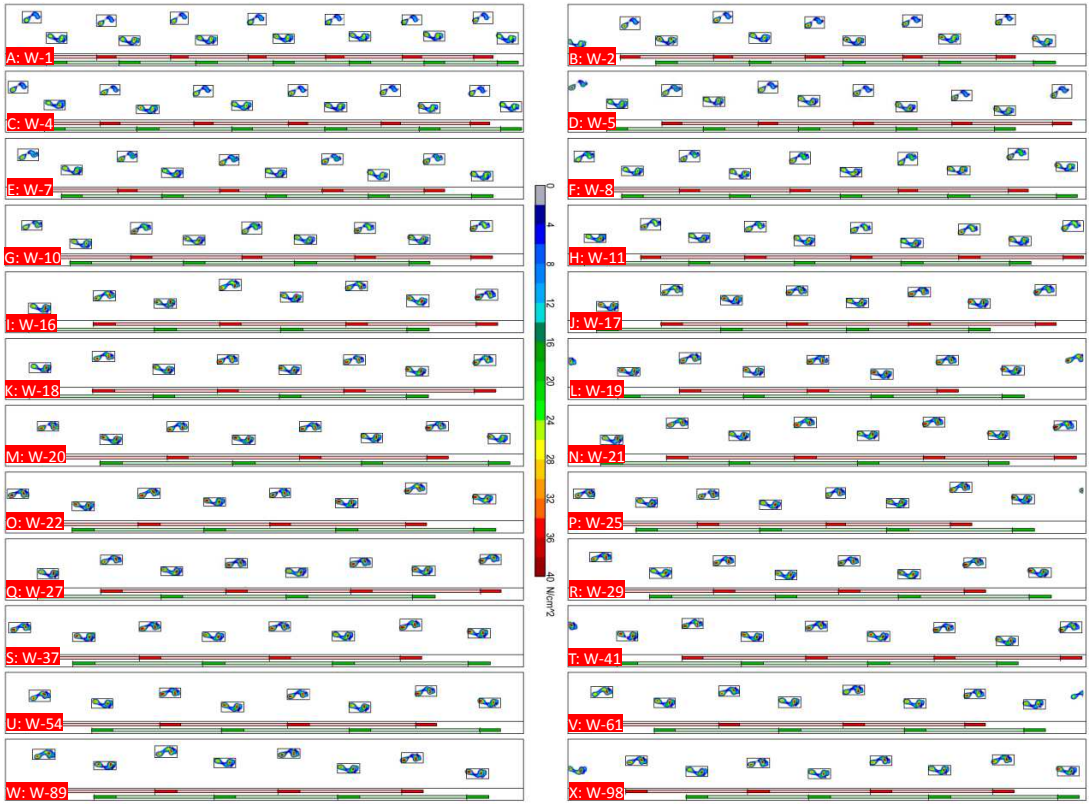}
\end{tabular}
\caption{\label{fig2} Separate footprints. A - X were results from 24 tests, respectively. The W-number referred to the testing week. The red bar stood for the left foot print length, the green the right footprint length, and the framed footprints the valid ones, which were included in gait parameter calculation. The first datum of a footprint was the location where the heel contacted the ground, and the second one was the time when the heel contacted the ground.}
\end{center}
\end{figure}

The test equipment was Zebris FDM System - Gait Analysis (Long platform). Gait parameters were obtained from WinFDM report (export to ASCII), including step length, stance phase, double support, single support, stride length, cadence and speed. Footprint results were shown in Fig.~\ref{fig2}.

\begin{figure}[!ht]
\begin{center}
\begin{tabular}{cccc}
 \includegraphics[width=14.8cm]{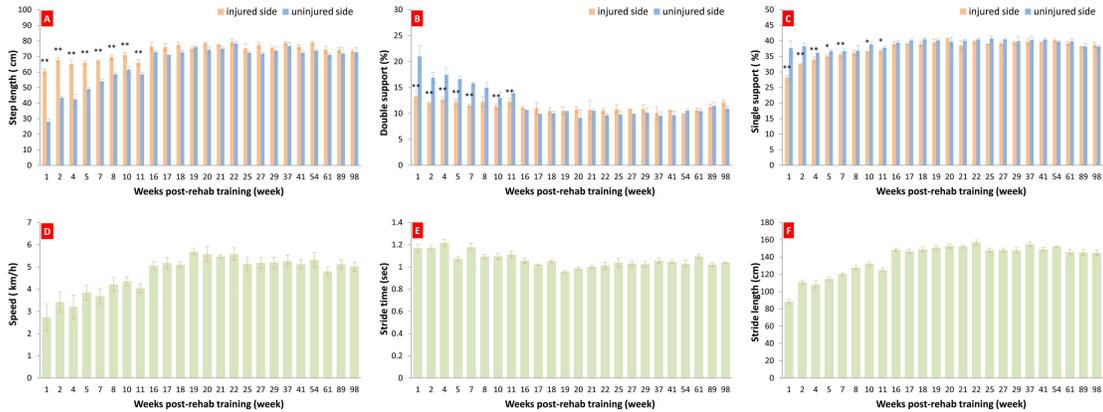}
\end{tabular}
\caption{\label{fig3} Gait parameters. A Step length. B Double support. C Single support. D Speed. E Stride time. F Stride length. The definitions of these concepts were retrieved from literature (). In (A) - (E), $*$ stood for $p<0.05$, and $**$ for $p<0.01$. T-TEST used the two-tailed distribution, and two-sample unequal variance (heteroscedastic).}
\end{center}
\end{figure}

Fig.~\ref{fig2} showed that in 24 tests, the stride length of both sides was almost the same. Speed was the result of stride length dividing stride time, so the stride time was almost the same. This verified Eq.~(\ref{eqn-2}) and Eq.~(\ref{eqn-3}). The tested gait parameters were shown in Fig.~\ref{fig3}.

\begin{figure}[!ht]
\begin{center}
\begin{tabular}{cccc}
 \includegraphics[width=14.8cm]{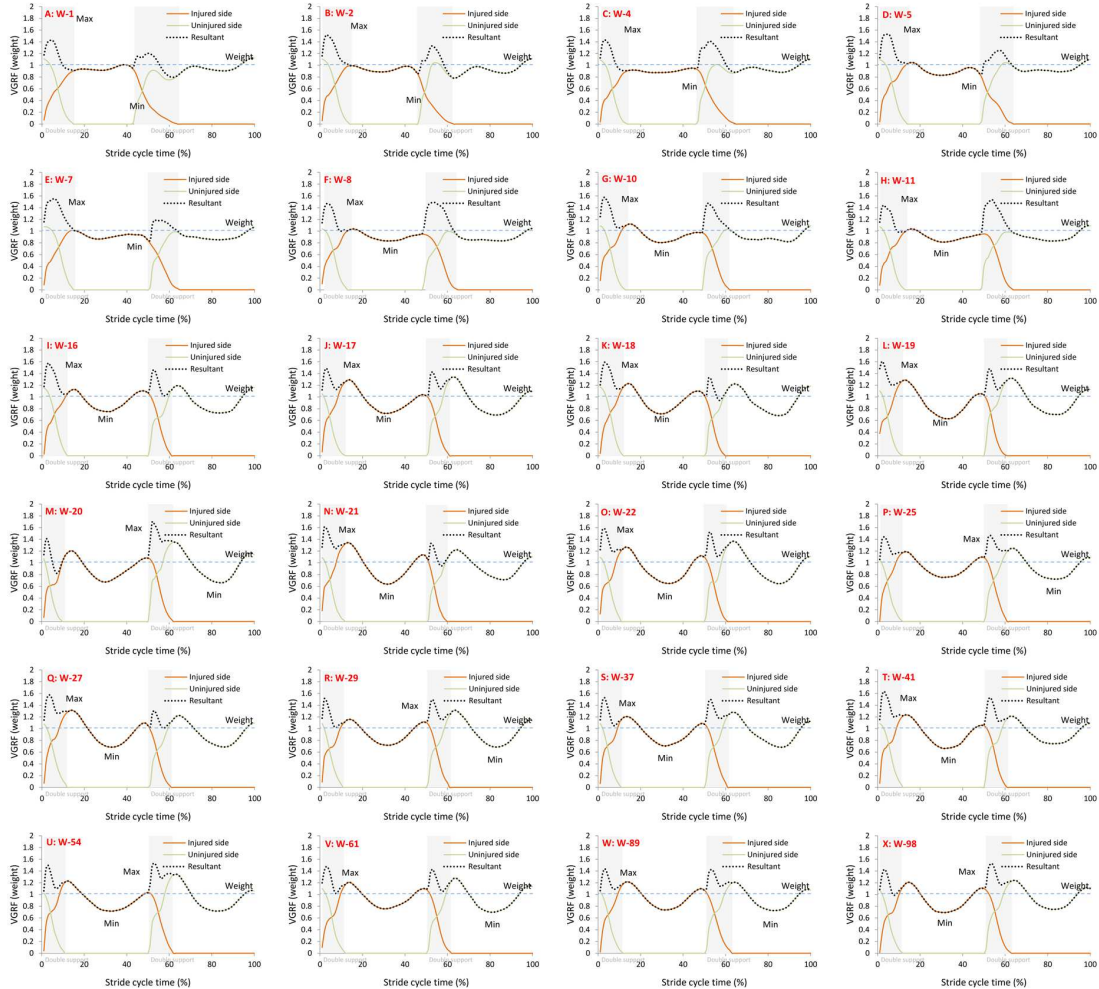}
\end{tabular}
\caption{\label{fig4} VGRF. A - X were results from 24 tests, respectively. The W-number referred to the testing week. Basic gait parameters were retrieved from WinFDM ($v1.2.2$) report (export to ASCII). Red stood for VGRF of the injured side, green VGRF of the uninjured side, and black square dot line the resultant force of both sides. Gray bar stood for double support time, and blue dash line for the weight. VGRF were by normalized by the weight.}
\end{center}
\end{figure}

Fig.~\ref{fig3} demonstrated that at the early stage of rehabilitation, the step length of the uninjured side was shorter than that of the injured side and significant difference was found. The double support and single support of the uninjured side were greater than those of the injured side, and significant difference was found. The speed was slowed down by reducing stride length. This verified Eq.~(\ref{eqn-4}), which was consistent with our gait pattern design. Could such gait pattern reflect the peak values of VGRF of the injured side? VGRF was thus tested. See Fig.~\ref{fig4}.

Fig.~\ref{fig4} showed that at the early stage of rehabilitation training, speed was slowed down by reducing stride length, the stance phase of the uninjured side was delayed and the step length of the uninjured side was reduced. The result was that the peak values of the injured side's VGRF were close to the weight.

Fig.~\ref{fig2}-~\ref{fig4} verified our hypothesis, indicating that Eq.~(\ref{eqn-1})-~\ref{fig4} presented biomechanical conditions of walking. The gait pattern for early rehabilitation traing of lower limb injury was designed on the scientific theory of Eq.~(\ref{eqn-1})-~\ref{fig4}.

\section{Conclusion}
To design a reasonable walking rehabilitation training program, conditions from the above four equations should be met. The simulation of these equations verified our hypothesis to reduce VGRF of the injured side to be close to the weight by delaying the uninjured side's stance phase, and by reducing the uninjured side's step length. Furthermore, based upon the relation between speed, cadence and stride length, the design of gait pattern of "reducing stride length to reduce speed, increasing the uninjured side's stance phase and reducing the uninjured side's step length" worked effectively. This also verified our hypothesis that when walking, by changing gait pattern, the injured side's VGRF peak values could be reduced to be close to the weight. It is advisable to design a gait pattern of lower limb injury rehabilitation training program according to these biomechanical conditions.

\section*{Acknowledgments}
This project was funded by the National Natural Science Foundation of China under the Grant Number $11172073$. The authors would like to acknowledge the support from the participant.

\end{document}